\documentclass[twocolumn,showpacs,amsmath,amssymb,nofootinbib,superscriptaddress]{
revtex4}

\usepackage{graphicx}
\usepackage{epstopdf}
\usepackage{epsfig}
\usepackage{bm}
\usepackage{amsfonts}
\usepackage[latin1]{inputenc}
\usepackage{amssymb}
\usepackage{color}
\usepackage{float}
\usepackage{mathtools}
\usepackage{amsmath}                
\usepackage{dcolumn}
\usepackage{hhline}
\usepackage{hyperref}
\usepackage{array}
\usepackage{lscape}
\usepackage{array}
\usepackage{booktabs}
\usepackage{multirow}

\providecommand{\U}[1]{\protect\rule{.1in}{.1in}}

\newcommand{\newc}{\newcommand}

\newc{\be}{\begin{equation}}
\newc{\ee}{\end{equation}}

\newc{\ba}{\begin{eqnarray}}
\newc{\ea}{\end{eqnarray}}
\newc{\bea}{\begin{eqnarray*}}
\newc{\eea}{\end{eqnarray*}}
\newc{\D}{\partial}
\newc{\ie}{{\it i.e.} }
\newc{\eg}{{\it e.g.} }
\newc{\etc}{{\it etc.} }
\newc{\etal}{{\it et al.}}
\newc{\lcdm}{$\Lambda$CDM }

\newc{\ra}{\Rightarrow}
 
\allowdisplaybreaks

\begin{document}

\title{Modified cosmology through nonextensive horizon thermodynamics }

\author{Andreas Lymperis}\email{alymperis@upatras.gr} 
\affiliation{Department of Physics, University of Patras, 26500 Patras, Greece}

\author{Emmanuel N. Saridakis}
\email{Emmanuel\_Saridakis@baylor.edu}
\affiliation{Department of Physics, National Technical University of Athens, Zografou
Campus GR 157 73, Athens, Greece}
\affiliation{CASPER, Physics Department, Baylor University, Waco, TX 76798-7310, USA}


%%%%%%%%%%%%%%
\begin{abstract}
We construct  modified cosmological scenarios through the application of 
the first law of thermodynamics on the universe horizon, but using the generalized, 
nonextensive Tsallis entropy instead of the usual Bekenstein-Hawking one. We result 
to modified cosmological equations that possess the usual ones as a particular limit, but 
which in the general case contain extra terms that appear for the first time, that  
constitute an effective dark energy sector quantified by the nonextensive parameter 
$\delta$. When the matter sector is dust, we extract analytical expressions for the dark 
energy density and equation-of-state parameters, and we extend these solutions to the 
case where radiation is present too. We show that the universe exhibits the 
usual thermal history, with the sequence of matter and dark-energy eras,  and according 
to the value of $\delta$ the dark-energy equation-of-state parameter  can be 
quintessence-like, phantom-like, or experience the phantom-divide crossing during the 
evolution. Even in the case where  the explicit cosmological constant is absent, the 
scenario at hand can very efficiently mimic $\Lambda$CDM cosmology, and is in excellent 
agreement with Supernovae type Ia observational data.
 \end{abstract}

\maketitle

\section{Introduction}

Recent cosmological observations from various and different fields reveal that the 
universe has experienced two accelerated expansion phases, one at early and one at late 
times. Since the established knowledge of general relativity and Standard 
Model of particles is not sufficient to explain this behavior, there has been a lot of 
effort in constructing theories beyond the above, in order to acquire the necessary extra 
degrees of freedom. On one hand, one can introduce new forms of matter, such as the 
inflaton field  \cite{Olive:1989nu,Bartolo:2004if} or the concept of dark energy   
\cite{Copeland:2006wr,Cai:2009zp}, which in the framework of general relativity can lead 
to the aforementioned accelerated behaviors. On the other hand, one can construct 
gravitational modifications, which possess general relativity as a particular limit, but 
at large scales can provide extra degrees of freedom capable of driving the 
acceleration (for reviews see \cite{Nojiri:2006ri,Capozziello:2011et,Cai:2015emx}). Note 
that this last approach has the additional theoretical advantage that may improve 
renormalizability, which seems to be necessary towards quantization 
\cite{Stelle:1976gc,Biswas:2011ar}. 
 
The usual approach of constructing modified gravitational theories is to start from the 
Einstein-Hilbert action and add correction terms. The simplest extension is to replace 
the Ricci scalar $R$ by a function $f(R)$   
\cite{Starobinsky:1980te,Capozziello:2002rd,DeFelice:2010aj,Nojiri:2010wj}. Similarly, one 
can 
proceed in constructing many other classes of modification, 
such as  $f(G)$ gravity \cite{Nojiri:2005jg, DeFelice:2008wz}, Lovelock 
gravity \cite{Lovelock:1971yv, Deruelle:1989fj}, Weyl gravity
\cite{Mannheim:1988dj, Flanagan:2006ra} and Galileon theory 
\cite{Nicolis:2008in, Deffayet:2009wt, Leon:2012mt}. Alternatively, one 
can start from the torsional formulation of gravity and build various extensions, such as 
$f(T)$ gravity  
\cite{Ben09, Linder:2010py, Chen:2010va}, $f(T,T_G)$ gravity
\cite{Kofinas:2014owa,Kofinas:2014daa}, etc.

On the other hand, there is a well-known conjecture that one can
express the Einstein equations as the first law of thermodynamics
\cite{Jacobson:1995ab,Padmanabhan:2003gd,Padmanabhan:2009vy}. In the particular case of 
cosmology in a universe filled with the matter and dark-energy fluids,
one can express the Friedmann equations as the first law of thermodynamics applied in the 
universe apparent 
horizon considered as a thermodynamical system   
\cite{Frolov:2002va,Cai:2005ra,Akbar:2006kj,Cai:2006rs}. Reversely, one can apply the 
first law of thermodynamics in the universe horizon, and extract 
the Friedmann equations. Although this procedure is a conjecture and not a proven 
theorem, 
it seems to work perfectly in a variety of modified gravities, as long as one uses the 
modified entropy relation that corresponds to each specific theory 
\cite{Cai:2006rs,Akbar:2006er,Paranjape:2006ca,Sheykhi:2007zp,Jamil:2009eb,Cai:2009ph,
Wang:2009zv,
Jamil:2010di, Gim:2014nba, Fan:2014ala}. Nevertheless, note 
that in order to know the modified entropy relation of a modified gravity, ones needs 
to know this modified gravity a priori and investigate it in spherically symmetric 
backgrounds.
In this sense the above procedure cannot provide new gravitational modifications, 
offering only a way to study their features.

In the present work we are interested in following the above procedure in a reverse way, 
in order to construct new cosmological modifications. In particular, we will apply the 
first law of thermodynamics, but instead of the usual entropy relation we will use the 
nonextensive, Tsallis entropy  
\cite{Tsallis:1987eu,Lyra:1998wz,Wilk:1999dr}, which is the consistent 
generalization of 
the Boltzmann-Gibbs additive entropy in non-additive  systems, such as gravitational 
ones. 
In this way we will obtain new modified Friedmann equations that possess the usual ones 
as 
a particular limit, namely when the Tsallis generalized entropy becomes the usual one, 
but which in the general case contain extra terms that appear for the first time. Hence, 
we will investigate in detail the cosmological implications of these new extra terms.

The plan of the work in the following: In Section \ref{Themodel} we present the 
construction of the scenario, applying the  first law of thermodynamics in the 
universe horizon, but using the generalized, 
nonextensive Tsallis entropy instead of the usual Bekenstein-Hawking one. In Section 
\ref{CosmEvol} we investigate the cosmological evolution, focusing on the behavior of the 
  dark energy density and equation-of-state parameters, studying separately the cases 
where an explicit cosmological constant is present or absent. Finally, in 
Section \ref{Conclusion} we summarize our results.

%δες γεβεραλιζεδεντοπυ ανδ εχει εφαρμοστει

\section{The model}
\label{Themodel}

In this section we present the scenario at hand, namely we extract modified Friedmann 
equations applying the first law of thermodynamics to the whole universe, but using the 
generalized Tsallis entropy instead of the standard one. Throughout the work we consider 
a  homogeneous and isotropic Friedmann-Robertson-Walker (FRW) geometry with metric
\begin{equation}
ds^2=-dt^2+a^2(t)\left(\frac{dr^2}{1-kr^2}+r^2d\Omega^2 \right),
\label{metric}
\end{equation}
  where $a(t)$ is the scale factor, and with $k=0,+1,-1$ corresponding to flat, close 
and open spatial geometry respectively.

\subsection{Friedmann equations as the first law of thermodynamics}
\label{basicmodel}

Let us first briefly review the extraction of the Friedmann equations in the case 
of general relativity, from the application of the first law of thermodynamics. 
We start by considering the expanding universe filled with 
the matter perfect fluid, with energy density $\rho_m$ and pressure $p_m$. 
Although it is not 
trivial what it should be its ``radius'', namely the length that forms its boundary, 
there 
is a consensus that one should use the apparent horizon 
\cite{Frolov:2002va,Cai:2005ra,Cai:2008gw}
\begin{equation}
\label{apphor}
 \tilde{r}_a=\frac{1}{\sqrt{H^2+\frac{k}{a^2}}},
\end{equation}
with $H=\frac{\dot a}{a}$ the Hubble parameter and dots 
denoting derivatives with respect to $t$.
The   apparent horizon is a marginally trapped surface with vanishing expansion, 
defined in 
general by  
the expression $h^{ij}\partial_i\tilde r\partial _j
\tilde r=0 $ (which implies that the vector $\nabla \tilde r$ is
null or degenerate on the apparent horizon surface)
\cite{Bak:1999hd}. For a dynamical spacetime, the apparent horizon is a causal horizon 
associated with the gravitational entropy and the surface gravity
\cite{Bak:1999hd,Hayward:1997jp,Hayward:1998ee}. Finally, note that in flat spatial 
geometry the apparent horizon becomes the Hubble one.  

The crucial point in the application of thermodynamics in cosmology is that the first law 
is interpreted in terms of energy flux and area of local Rindler horizons, and that heat 
is defined as energy that flows across a causal horizon,
and hence thermodynamics is applied on the horizon itself, considered as a system 
separated not by  a diathermic wall but by a causality barrier
\cite{Jacobson:1995ab,Padmanabhan:2003gd,Padmanabhan:2009vy}. One can 
attribute to the universe horizon a temperature and an entropy that arise from the 
corresponding relations of black hole temperature and entropy respectively, but with the 
universe horizon, namely the apparent horizon, in place of the black hole horizon.
Concerning the black hole temperature, it is well known that for spherically symmetric 
geometry its relation does not depend on the underlying gravitational theory, and it is 
just inversely proportional to the black hole horizon, namely 
$T=1/(2\pi r_h)$ \cite{Gibbons:1977mu}. Hence, one can attribute to the universe horizon 
the temperature  \cite{Padmanabhan:2009vy}
\begin{equation}
\label{Th}
 T_h=\frac{1}{2\pi\tilde{r}_a},
\end{equation} 
independently of the gravitational theory that governs the universe. Concerning the 
back hole entropy, it is also known that its relation does depend on the underlying 
gravitational theory \cite{Padmanabhan:2009vy}. In the case of general relativity one 
obtains the usual Bekenstein-Hawking relation $S=A/(4G)$ (in units where $\hbar=k_B = c 
= 1$), where $A=4\pi r_h^2$ is the area of the black hole and $G$ the gravitational 
constant. Thus, in the case of a universe governed by general relativity, the horizon 
entropy will be just 
\begin{equation}
\label{Horentropy}
S_h=\frac{1}{4G} A.
\end{equation}
Finally, a last reasonable assumption is that after equilibrium establishes 
the universe fluid acquires the same temperature with the horizon one,
otherwise the energy flow would deform this geometry 
\cite{Izquierdo:2005ku}.\footnote{Note that although this will certainly be the 
situation at late times,  when the universe fluid and the horizon will have
interacted for a long time, it is not assured that it will be the case  
at early or intermediate times. However, in order to avoid applying non-equilibrium 
thermodynamics, which leads to mathematical complexity, the assumption of 
equilibrium  is widely used 
\cite{Padmanabhan:2009vy,Frolov:2002va,Cai:2005ra,Akbar:2006kj,Izquierdo:2005ku,
Jamil:2010di}.
Thus, we will follow this assumption and we will have in mind that
our results hold only at late times of the universe evolution.}

As the universe evolves an amount of energy from the universe fluid crosses the horizon. 
During an
infinitesimal time interval $dt$, the heat flow that crosses the horizon  can be 
straightforwardly found to be  \cite{Cai:2005ra}
 \be \label{energy}
\delta Q=-dE=A(\rho_m+p_m)H \tilde{r_{a}}dt,
\ee
 with $A=4\pi r_{a}^2$ the apparent horizon area. On the other hand, the first law of 
thermodynamics states that $-dE=TdS$. Since the temperature and entropy of the horizon 
are given by (\ref{Th}) and (\ref{Horentropy}) respectively, we find that $dS=2\pi 
\dot{\tilde{r}}_a dt/G$, with $\dot{\tilde{r}}_a$ easily obtained from (\ref{apphor}). 
Inserting the above into the first law of thermodynamics we finally acquire 
\be
\label{cFE1}
-4\pi G (\rho_m +p_m)= \dot{H} - \frac{k}{a^2}.
\ee 
Additionally, assuming that the matter fluid satisfies the conservation equation
\be
\label{consrvationequation}
 \dot{\rho}_m +3H(\rho_m +p_m)=0,
 \ee 
 inserting it into (\ref{cFE1}) and integrating we obtain 
\be 
\label{cFE2}
\frac{8\pi G}{3}\rho_m =H^2+\frac{k}{a^2}-\frac{\Lambda}{3},
\ee
with $\Lambda$ the integration constant, that plays the role of a cosmological constant.

Interestingly enough, we saw that applying the first law of thermodynamics to the whole 
universe resulted to the extraction of the two Friedmann equations, namely Eqs. 
(\ref{cFE1}) and (\ref{cFE2}). The above procedure can be extended to modified gravity 
theories too, where as we discussed the only change will be that the entropy relation 
will not be the general relativity one, namely (\ref{Horentropy}), but the one 
corresponding to the specific modified gravity at hand 
\cite{Cai:2006rs,Akbar:2006er,Paranjape:2006ca,Sheykhi:2007zp,Jamil:2009eb,Cai:2009ph,
Wang:2009zv,
Jamil:2010di, Gim:2014nba, Fan:2014ala}. Nevertheless, we have to mention here
 that although the above procedure offers a significant tool to study the features and 
properties of   various modified gravities, it does not lead to  new gravitational 
modifications, since one needs to know the entropy relation, which in turn can be known 
only if a specific modified gravity is given a priori.

\subsection{ Tsallis entropy}
\label{Tsallisentropy}

In this subsection we briefly review the concept of nonextensive, or Tsallis entropy 
\cite{Tsallis:1987eu,Lyra:1998wz,Wilk:1999dr}. 
As Gibbs pointed 
out already at 1902, in systems where the partition function diverges, the standard  
Boltzmann-Gibbs theory is not applicable, and large-scale gravitational systems are 
known to fall within this class. Tsallis generalized standard thermodynamics (which  
arises from the hypothesis of weak probabilistic correlations and their connection to 
ergodicity) to  
nonextensive one, which can be applied in all cases, and still possessing standard 
Boltzmann-Gibbs theory as a limit. Hence, the usual 
Boltzmann-Gibbs additive entropy must be generalized to the nonextensive, i.e 
non-additive entropy (the entropy of the 
whole system is not necessarily the sum of the entropies of its sub-systems), which is 
named Tsallis entropy  
\cite{Tsallis:1987eu,Lyra:1998wz,Wilk:1999dr,Nunes:2014jra,Saridakis:2018unr}. In cases of 
spherically 
symmetric systems 
that we are interested in this work, it can be written  in compact form as   
\cite{Tsallis:2012js}:
\begin{equation}
\label{Tsalsent}
S_T=\frac{\tilde{\alpha}}{4G} A^{\delta}, 
\end{equation}
in units where $\hbar=k_B = c = 1$,
where $A\propto L^2$ is the area of the system with characteristic length $L$, $G$ is the 
gravitational constant, $\tilde{\alpha}$ is a positive constant with dimensions 
$[L^{2(1-\delta)}]$ and $\delta$ 
denotes the non-additivity parameter.\footnote{In \cite{Tsallis:2012js} the nonextensive 
entropy relation is written as $S_T=\gamma A^\delta$, however we prefer to write is as 
in (\ref{Tsalsent}) for convenience.} Under the hypothesis of equal 
probabilities the parameters $\delta$  and $\tilde{\alpha}$ are related to the 
dimensionality of the system  \cite{Tsallis:2012js} (in particular the important parameter
$\delta=d/(d-1)$ for $d>1$), however in the general case they remain  independent and 
free 
parameters. Obviously, in the 
case $\delta=1$ and $\tilde{\alpha}=1$, Tsallis entropy becomes the usual  
Bekenstein-Hawking additive
entropy.

\subsection{Modified Friedmann equations through nonextensive  first law of 
thermodynamics}
\label{fullmodelconstr}

In subsection \ref{basicmodel} we presented the procedure to extract the Friedmann 
equations from the first law of thermodynamics. This procedure can be applied in any 
modified gravity, as long as one knows the black hole entropy relation for this specific 
modified gravity. Hence, as we mentioned above,  although it can be enlightening for the 
properties of various modified gravities, the thermodynamical approach   does not lead to 
 new gravitational modifications since one needs to consider a specific 
modified gravity a priori.

In the present subsection however, we desire to follow the steps of subsection 
\ref{basicmodel}, but instead of the standard additive entropy relation to use the 
generalized, nonextensive, Tsallis entropy presented in subsection \ref{Tsallisentropy} 
above. Doing so we do obtain modified Friedmann equations, with modification terms 
that appear for the first time, and which provide the standard Friedmann equations in the 
case where Tsallis entropy becomes the standard Bekenstein-Hawking one.

We start from the first law of thermodynamics $-dE=TdS$, where $-dE$ is given by 
(\ref{energy}), $T$ by (\ref{Th}), but we will consider that the entropy is given by 
Tsallis entropy (\ref{Tsalsent}). In this case, and recalling that $A=4\pi \tilde{r}_a^2$
we acquire
\begin{equation}
 dS=(4\pi)^\delta\frac{\delta\tilde{\alpha}}{2G}\tilde{r_{a}}^{2\delta-1}\dot{\tilde{r}}
_a dt.
\end{equation}
Inserting everything in the first law, and calculating $\dot{\tilde{r}}_a$ from 
(\ref{apphor}), we obtain
\be \label{gfe1}
-\frac{(4\pi)^{2-\delta}G}{\tilde{\alpha}}(\rho_m+p_m)=\delta 
\frac{\dot{H}-\frac{k}{a^2}}{\left(H^2+\frac{k}{
a^2}\right)^{\delta -1}}.
\ee
Finally, inserting the conservation equation (\ref{consrvationequation}) and integrating, 
for $\delta\neq2$ we obtain
\be \label{gfe2}
\frac{2(4\pi)^{2-\delta}G}{3\tilde{\alpha}} \rho_m=\frac{ \delta 
}{2-\delta} \left(H^2+\frac{k}{a^2}\right) 
^{2-\delta}-\frac{\tilde{\Lambda}}{3\tilde{\alpha}},
\ee
where $\tilde{\Lambda}$ is an integration constant. Hence, the use of Tsallis entropy in 
the first law of thermodynamics, led to two modified Friedmann equations, namely 
(\ref{gfe1}) and (\ref{gfe2}), with modification terms that appear for the first time 
depending on three parameters out of which two are free.

Let us elaborate the obtained modified Friedmann equations. From now on we focus on the 
flat 
case, namely we consider $k=0$, which allows us to extract analytical expressions, 
however the investigation of the non-flat case is straightforward.
We can re-write  (\ref{gfe1}), (\ref{gfe2}) as 
\begin{eqnarray}
\label{FR1}
&&H^2=\frac{8\pi G}{3}\left(\rho_m+\rho_{DE}\right)\\
&&\dot{H}=-4\pi G \left(\rho_m+p_m+\rho_{DE}+p_{DE}\right),
\label{FR2}
\end{eqnarray}
where we have defined the effective dark energy density and pressure as 
\begin{eqnarray}
&&
\!\!\!\!\!\!\!\!
\rho_{DE}=\frac{3}{8\pi G} 
\left\{(4\pi)^{\delta-1}\frac{\tilde{\Lambda}}{3}\right.\nonumber\\
&&\left.\ \ \  \ \ \ \ \ \ \ \ \ \ \ \ 
+H^2\left[1-\tilde{\alpha}(4\pi)^{
\delta-1}\frac{ \delta}{ 2\!-\!\delta} H^{2(1-\delta) }
\right]
\right\},
\label{rhoDE}\\
&& 
\!\!\!\!\!\!\!\!
p_{DE}= -\frac{1}{8\pi G}\left\{
(4\pi)^{\delta-1} 
\tilde{\Lambda} +2\dot{H}\left[1-\tilde{\alpha} (4\pi)^{\delta-1}\delta H^{2(1-\delta)}
\right]
\right.\nonumber\\
&&\left.\ \ \  \ \ \ \ \ \ \ \ \ \ \ \ 
+3H^2\left[1-\tilde{\alpha}(4\pi)^{\delta-1}\frac{\delta}{2\!-\!\delta}H^{
2(1-\delta)}
\right]
\right\}.
\label{pDE}
\end{eqnarray}
We can further simplify the above expressions by redefining 
$ \Lambda  \equiv (4\pi)^{\delta-1}\tilde{\Lambda}$ and 
$\alpha\equiv (4\pi)^{\delta-1}\tilde{\alpha}$, obtaining
\begin{eqnarray}
&&
\!\!\!\!\!\!\!\!\!\!\!\!\!\!\!\!\!\!
\rho_{DE}=\frac{3}{8\pi G} 
\left\{ \frac{\Lambda}{3}+H^2\left[1-\alpha \frac{ \delta}{ 2-\delta} 
H^{2(1-\delta) }
\right]
\right\},
\label{rhoDE1}
\end{eqnarray}
\begin{eqnarray}
&& \!\!\!\!\!\!\!\!\!\!\!\!\!\!\!\!\!\!\!\!
p_{DE}= -\frac{1}{8\pi G}\left\{
\Lambda
+2\dot{H}\left[1-\alpha\delta H^{2(1-\delta)}
\right]\right.\nonumber\\
&&\left. \ \ \ \ \ \ \ \ \ \ 
+3H^2\left[1-\alpha\frac{\delta}{2-\delta}H^{
2(1-\delta)}
\right]
\right\}.
\label{pDE1}
\end{eqnarray}
Thus, we can define the equation-of-state parameter for the effective dark energy 
sector 
as
\begin{eqnarray}
w_{DE}\equiv\frac{p_{DE}}{\rho_{DE}}=-1-
\frac{     
  2\dot{H}\left[1-\alpha\delta H^{2(1-\delta)}
\right]
 }{\Lambda+3H^2\left[1-\frac{\alpha\delta}{2-\delta}H^{2(1-\delta)}
\right]}
\label{wDE}.
\end{eqnarray}

In summary, in the constructed  modified cosmological scenario, equations 
(\ref{consrvationequation}), (\ref{FR1}) and (\ref{FR2}) can determine the universe 
evolution, as long as the matter equation-of-state parameter is known. In particular, 
inserting (\ref{rhoDE1}), (\ref{pDE1}) 
into (\ref{FR2}), we acquire a differential
equation for $H(t)$ that can be solved similarly to all 
modified-gravity and dark-energy models.

Finally, as one can see, in the case $\delta=1$ and $\alpha=1$ the generalized 
Friedmann equations  (\ref{FR1}),(\ref{FR2}) reduce to $\Lambda$CDM cosmology, namely 
\begin{eqnarray}
&&H^2=\frac{8\pi G}{3} \rho_m+\frac{\Lambda}{3}\nonumber\\
&&\dot{H}=-4\pi G(\rho_m+p_m).
\end{eqnarray}

We close this subsection by providing for completeness the equations for
$\delta=2$. In this special case, integration of (\ref{gfe1}), instead of (\ref{gfe2})
results to 
\be \label{soldelta2}
\frac{G}{3 \tilde{\alpha}} \rho_m=\ln\left[H^2+\frac{k}{a^2}
\right] -\frac{\tilde{\Lambda}}{6\tilde{\alpha}}.
\ee 
Hence, in this case the two Friedmann equations (\ref{gfe1}) and (\ref{soldelta2}), for 
$k=0$,
lead to the definitions
{\begin{eqnarray}
&& \!\!\!\!\!\!\!\!\!\!\!\!\!\!\!
\rho_{DE}=\frac{3}{8\pi G} 
\left[\frac{ \Lambda}{3}+H^2-2\alpha\ln H^2
\right]
\label{rhoDEd2}\\
&& \!\!\!\!\!\!\!\!\!\!\!\!\!\!\!
p_{DE}= -\frac{1}{8\pi G}\left[
 \Lambda\!+\!3H^2\! -\!6\alpha\ln H^2
 \!+\!2\dot{H}\left(1\!-\!   \frac {2\alpha}{H^2} \right)\!
\right]\!,
\label{pDEd2}
\end{eqnarray}

and thus
\begin{eqnarray}
w_{DE}\equiv\frac{p_{DE}}{\rho_{DE}}=-1-
\frac{  
 2\dot{H}\left(1-   \frac {2\alpha}{H^2} \right)
 }{\Lambda+3H^2-6\alpha\ln H^2}.
\end{eqnarray}

\section{Cosmological evolution}
\label{CosmEvol}

In this section we proceed to a detailed investigation of the modified cosmological 
scenarios constructed above. The cosmological equations are the two modified Friedmann 
equations  (\ref{FR1}) and (\ref{FR2}), along with the conservation 
equation (\ref{consrvationequation}). In the general case of a general matter 
equation-of-state parameter, $w_m\equiv p_m/\rho_m$, analytical solutions cannot be 
extracted, and thus one has to solve the above equations numerically. However, we are 
interested in providing analytical expressions too, and thus in the following we focus to 
the case of dust matter, namely $w_m=0$.

As usual for convenience we introduce the matter and dark energy density parameters 
respectively as
 \begin{eqnarray} \label{omatter}
&&\Omega_m=\frac{8\pi G}{3H^2} \rho_m\\
&& \label{ode}
\Omega_{DE}=\frac{8\pi G}{3H^2} \rho_{DE}.
 \end{eqnarray} 
In the case of dust matter,  equation (\ref{consrvationequation}) gives that 
 $\rho_{m} = \frac{\rho_{m0}}{a^3}$, with $\rho_{m0}$ the value of the matter energy 
density at present scale factor $a_0=1$ (in the following the subscript ``0" marks the 
present value of a quantity). Therefore, in this case equation (\ref{omatter})   gives 
immediately 
$\Omega_m=\Omega_
{m0} H_{0}
^2/a^3 H^2$. Combining this with the fact that $\Omega_m + \Omega_{DE}=1$ we can 
easily extract that
\be \label{h2}
H=\frac{\sqrt{\Omega_{m0}} H_{0}}{\sqrt{a^3 (1-\Omega_{DE})}}.
\ee
In the following we will use the redshift $z$ as the independent variable, defined  as  $ 
 1+z=1/a$ for $a_0=1$. Thus, differentiating (\ref{h2}) we can obtain the useful 
expression
\be \label{hddot}
\dot H=-\frac{H^2}{2(1-\Omega_{DE})}[3(1-\Omega_{DE})+(1+z)\Omega'_{DE}],
\ee
where a prime denotes derivative with respect to $z$. 

Inserting (\ref{rhoDE1}) into (\ref{ode}) and using (\ref{h2}) we obtain 
 \begin{eqnarray} 
 \label{omegaDE}
&&
\!\!\!\!\!\!\!\!\!\!\!\!\!\!
\Omega_{DE}(z)=
1-H^{2}_{0}\Omega_{m0}(1+z)^3\nonumber\\
&&\ \ \ \ \ \ \ \ \, 
\cdot\left\{\frac{
(2\!-\!\delta)}{\alpha 
\delta}\left[H^{2}_{0}\Omega_{m0}(1\!+\!z)^3+\frac{\Lambda}{3}
\right]\right\}^{\frac{1}{\delta -2}}\!.
 \end{eqnarray}
This expression is the analytical solution for the dark energy density parameter  
$\Omega_{DE}(z)$, in a flat universe and for dust matter. Applying it at present time, 
i.e at $z=0$, we acquire
\begin{eqnarray} \label{lambda}
 \Lambda=\frac{3\alpha\delta}{2-\delta}H_0^{2(2-\delta)}-3H_0^2\Omega_{m0},
\end{eqnarray}
 which provides the relation that relates $\Lambda$, $\delta$ and $\alpha$ with the
observationally 
determined quantities $\Omega_{m0}$ and $H_0$, leaving the scenario with two free 
parameters. As expected, for $\delta=1$ and $\alpha=1$
all the above relations give those of $\Lambda$CDM cosmology.\\ \\

Differentiating (\ref{omegaDE}) we find
 \begin{eqnarray} 
\label{omegaDEdot}
&&\!\!\!\!\!\!\!\!\!\!\!\!\!\!\!\!
\Omega'_{DE}(z)=\left\{ 
\frac{(2-\delta)}{\alpha\delta}\left[1+\frac{\Lambda}{3}\frac{1}{\Omega_{m0}H^{2}_{0}(
1+z)^3}\right]\right\}^{\frac{3-\delta}{\delta -2}}
\nonumber\\
&& \ \ \,   \ 
\cdot
\frac{1}{\alpha 
\delta}\left[\Omega_{m0}H^{2}_{
0}(1+z)^3\right]^\frac{1}{\delta -2}\nonumber\\
&&
 \ \ \ \, 
 \cdot
\left[3(\delta -1)\Omega_{m0}H^{2}_{0}(1+z)^2+(\delta 
-2)\frac{\Lambda}{1+z}\right].
 \end{eqnarray}
 Hence, we can now calculate  the other important observable, namely the dark-energy 
equation-of-state parameter $w_{DE}$ from  (\ref{wDE}), eliminating $\dot{H}$ through 
(\ref{hddot}), obtaining
\begin{widetext}
\be
\label{wDEfinal}
w_{DE}(z)=-1+\frac{\left\{3[1-\Omega_{DE}(z)]+(1+z)\Omega_{DE}'(z)\right\}\left\{1-\alpha 
\delta 
\left[\frac{
H^{2}_{0}\Omega_{m0}(1+z)^3}{1-\Omega_{DE}(z)}\right]^{1-\delta}\right\}}{[1-\Omega_{DE}
(z)]\left\{\frac{\Lambda 
[1-\Omega_{DE}(z)]}{H^{2}_{0}\Omega_{m0}(1+z)^3}+3\left\{1-\frac{\alpha 
\delta}{2-\delta}\left[\frac{H^{2}_{0}\Omega_{m0}(1+z)^3}{1-\Omega_{DE}(z)}\right]^{ 
1-\delta}\right\} \right\}},
\ee
\end{widetext}
  where $\Omega_{DE}$ and $\Omega_{DE}'$ are given by (\ref{omegaDE}) and 
(\ref{omegaDEdot}) respectively. 
Lastly, it proves convenient to introduce the decelaration parameter $q\equiv 
-1-\frac{\dot H}{H^2}$, where using  (\ref{hddot}) is found to be 
\be \label{qpar}
q(z)=-1+\frac{1}{2[1-\Omega_{DE}(z)]}\{3[1-\Omega_{DE}(z)]+(1+z)\Omega'_{DE}(z)\}.
\ee

In summary, considering dust matter and flat geometry we were able to extract analytical 
solutions for $\Omega_{DE}(z)$ and $w_{DE}(z)$, for the modified, nonextensive 
cosmological scenarios of the present work. In the following two subsections we will 
investigate them in two distinct cases, 
namely when the explicit cosmological constant $\Lambda$ is present and when it is absent.

\subsection{Cosmological evolution with $\Lambda\neq 0$}

We first examine the case where the explicit cosmological constant $\Lambda$ is present. 
In this case when $\delta=1$ and $\alpha=1$ we obtain 
$\Lambda$CDM cosmology, and thus we are interested in studying the role of the 
nonextensive parameter $\delta$ on the cosmological evolution.

We use relation (\ref{lambda}) in order to set the value of $\Lambda$ that corresponds to 
$\Omega_{m0}\approx0.3$ in agreement with observations \cite{Ade:2015xua}. Moreover, in 
order to investigate the pure effect of $\delta$, we set $\alpha$ to its standard value, 
namely $\alpha=1$ (although for $\delta=1$ the parameter $\alpha$ is dimensionless, as we 
mentioned for  $\delta\neq1$ it acquires dimensions $[L^{2(1-\delta)}]$ and for 
convenience we use units where $H_0=1$).
\begin{figure}[!h]
\centering
\includegraphics[width=5.9cm]{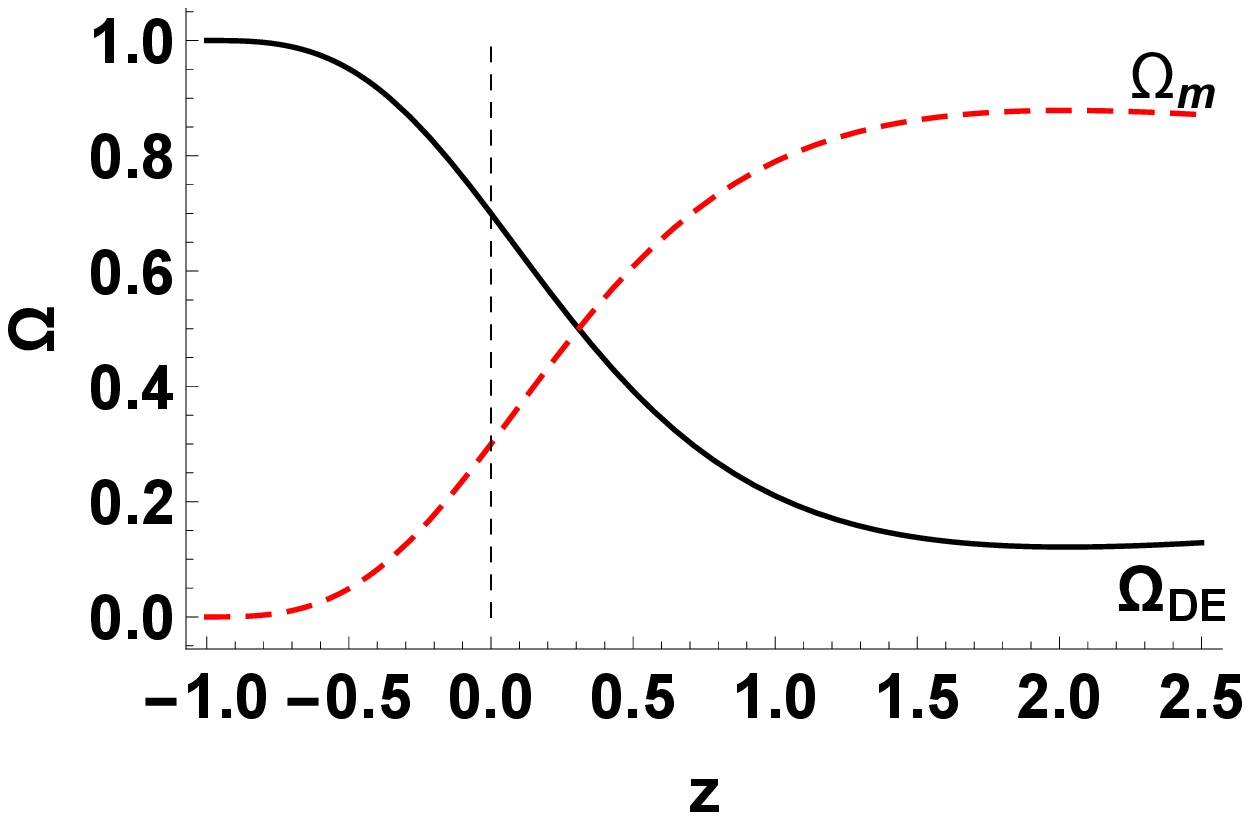} \\                                     
\includegraphics[width=5.9cm]{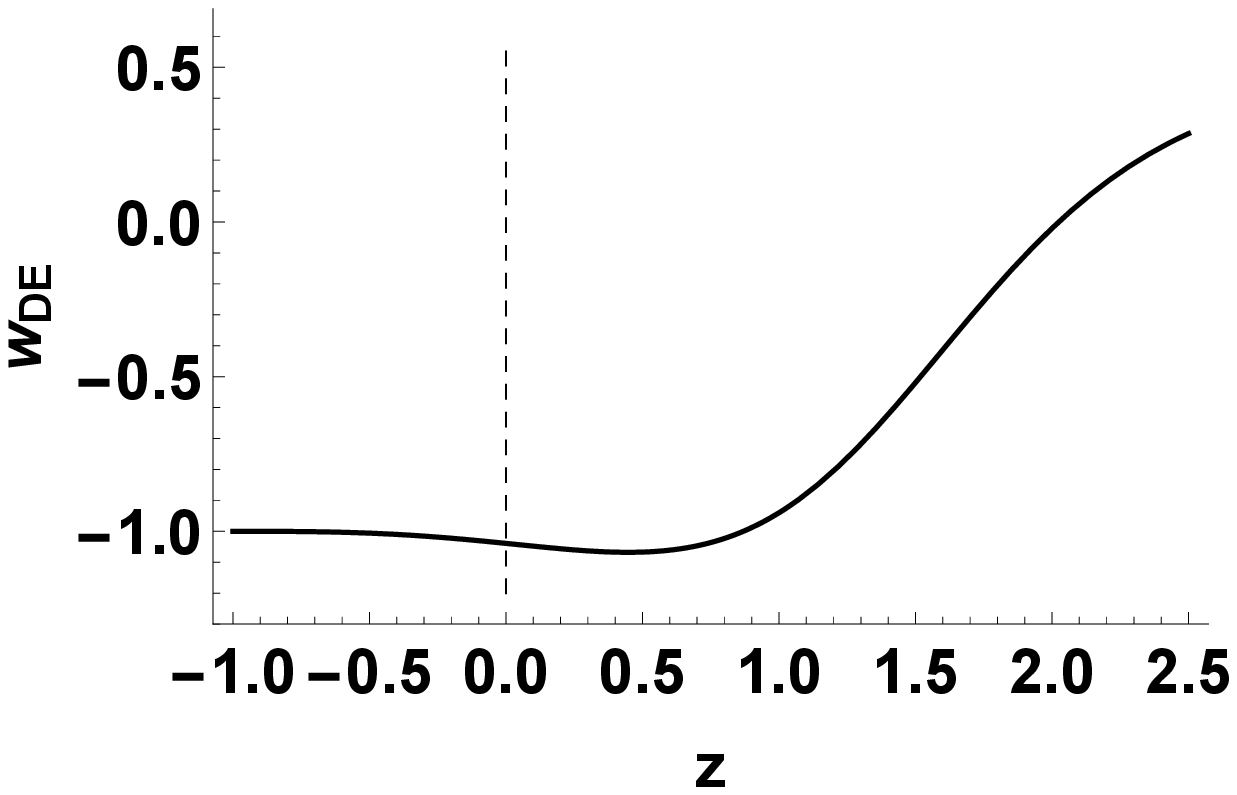} \\
\includegraphics[width=5.9cm]{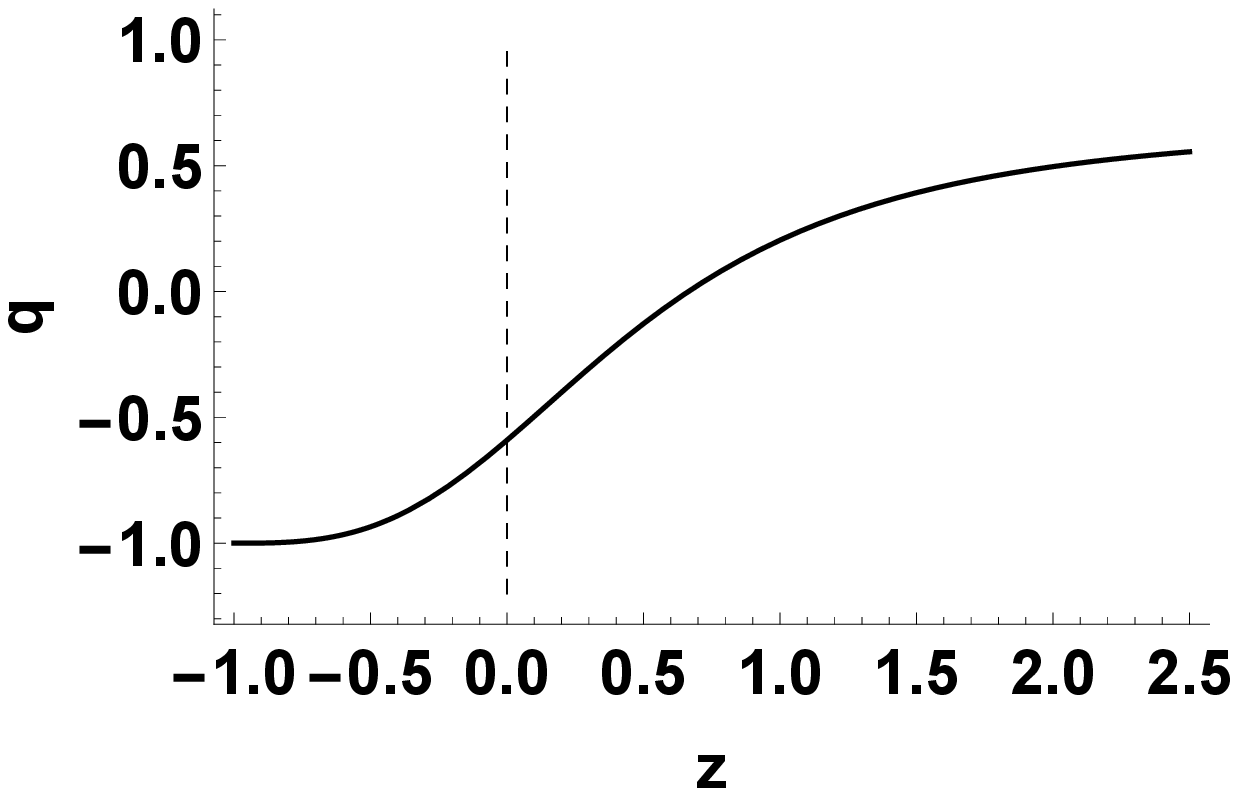}
\caption{\it{Upper graph: The evolution of the nonextensive dark energy density 
parameter $\Omega_{DE}$ (black-solid) and of the matter density parameter $\Omega_{m}$ 
(red-dashed), as a function of the redshift $z$, for $\delta=1.1$ and $\alpha=1$ in 
units where $H_0=1$. Middle graph: The evolution of the corresponding dark-energy 
equation-of-state parameter
$w_{DE}$. Lower graph: The evolution of the corresponding deceleration parameter $q$. In 
all graphs we have set the cosmological constant $\Lambda$ from
(\ref{lambda})  in order to obtain $\Omega_{m}(z=0)=\Omega_{m0}\approx0.3$ at present, 
and we have added a vertical dotted line denoting 
the present time $z=0$. 
}}
\label{fig:fig1}
\end{figure}
In the upper graph of Fig.  \ref{fig:fig1} we depict $\Omega_{DE}(z)$ and $\Omega_{m}(z) 
= 
1-\Omega_{DE}(z)$, as  given by equation 
(\ref{omegaDE}), in the case where $\delta=1.1$.
In the middle graph we present the corresponding evolution of $w_{DE}(z)$ 
according to (\ref{wDEfinal}). Finally, in the lower graph we present the deceleration 
parameter $q(z)$ from (\ref{qpar}). We mention that for transparency we have extended the 
evolution up to the far future, namely up to 
$z \rightarrow -1$, which corresponds to $t \to \infty$.

As we observe, we acquire the usual thermal history of the universe, with the sequence of 
matter and dark energy epochs, with the transition from deceleration to acceleration 
taking place at $z\approx 0.45$ in agreement with observations. Additionally, in 
the future the universe tends asymptotically to a complete dark-energy 
dominated, de-Sitter state. We mention the interesting bahavior that although at 
intermediate times the dark-energy equation-of-state parameter may experience the 
phantom-divide crossing and lie in the phantom regime, at asymptotically large times it 
will always stabilize at the cosmological constant value $-1$. Namely, the de-Sitter 
solution 
is a stable late-time attractor, which is a significant advantage (this can be easily 
showed taking the limit $z \rightarrow -1$ in (\ref{omegaDE}),(\ref{omegaDEdot}) and  
(\ref{wDEfinal}),  which gives $\Omega_{DE}\rightarrow 1$, $\Omega'_{DE}\rightarrow 0$, 
and  $w_{DE}\rightarrow-1$, respectively).

Let us now examine in detail the role of $\delta$ in the evolution, and in particular 
on $w_{DE}$. In Fig.  \ref{fig:fig2} we 
depict $w_{DE}(z)$ for $\alpha=1$ and for various values of $\delta$, including the value 
$\delta=1$ that reproduces $\Lambda$CDM cosmology. For each value of 
$\delta$ we  choose  $\Lambda$ according to
(\ref{lambda})  in order to obtain $\Omega_{m}(z=0)=\Omega_{m0}\approx0.3$ at present, 
and obtain an evolution of $\Omega_{DE}(z)$ and  $\Omega_m(z)$ similar to the upper graph 
of  
Fig.  \ref{fig:fig1}. In this way we can examine the pure effect of $\delta$. Firstly, as 
we mentioned, for $\delta=1$ we  obtain  $w_{DE}=-1=const.$, namely $\Lambda$CDM 
cosmology. For increasing $\delta>1$, at earlier redshifts $w_{DE}$ acquires larger 
values, while on the contrary in the recent past, i.e at $0\leq z\lesssim 0.8  
$,  $w_{DE}$ acquires algebraically  smaller values, which is also true for its present 
value $w_{DE0}$. In all cases the universe experiences the phantom-divide crossing, and 
in the far future it results from below in a de-Sitter phase with $w_{DE}$ being $-1$. On 
the other hand, for decreasing $\delta<1$ the behavior of  $w_{DE}(z)$ is the opposite, 
namely it initially lies in the phantom regime, it then crosses the $-1$-divide from  
below to above being quintessence-like at present, and finally it asymptotically tends to 
 $-1$ from above.
 \begin{figure}[!h]
\centering
\includegraphics[width=8.2cm]{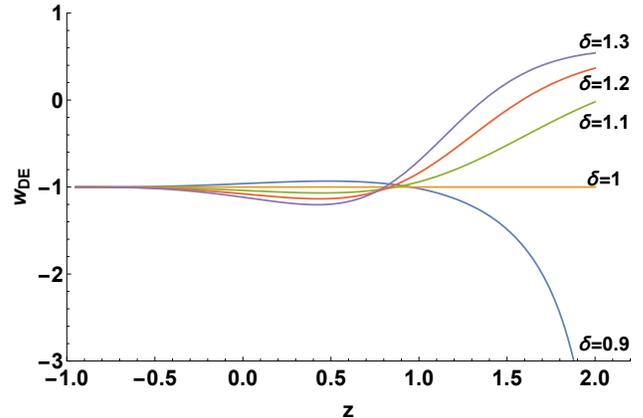}
\caption{\it{The evolution of the dark-energy equation-of-state parameter $w_{DE}$ as a 
function of the redshift $z$, for $\alpha=1$ in units where $H_0=1$, and various values 
of the nonextensive parameter $\delta$.  For each value of 
$\delta$ we  choose  $\Lambda$ according to
(\ref{lambda})  in order to obtain $\Omega_{m}(z=0)=\Omega_{m0}\approx0.3$ at present, 
and acquire an evolution of $\Omega_{DE}$ and  $\Omega_m$ similar to the upper graph of  
Fig.  \ref{fig:fig1}.}}
\label{fig:fig2}
\end{figure}

In summary, we can see that the nonextensive parameter $\delta$, that lies in the core of 
the modified cosmology obtained in this work, plays an important role in giving to dark 
energy a dynamical nature and bringing about a correction to $\Lambda$CDM 
cosmology. We mention that in all the above examples we kept the parameter $\alpha$ 
fixed, in order to maintain the one-parameter character of the scenario. Clearly, letting 
$\alpha$ vary too, increases the capabilities of the model and the obtained cosmological 
behaviors.

 \subsection{Cosmological evolution with $\Lambda=0$}
 \label{Lam0a}

In the previous subsection we investigated the scenario of modified Friedmann equations 
through nonextensive thermodynamics, in the case where the cosmological constant is 
explicitly present. Thus, we studied models that possess $\Lambda$CDM cosmology as a 
subcase, and in which the nonextensive parameter $\delta$ and its induced novel terms 
lead to corrections to  $\Lambda$CDM  paradigm.

In the present subsection we are 
interested in studying a more radical application of the scenario at hand, namely to 
consider that an explicit cosmological constant is not present and let the model 
parameters $\delta$ and $\alpha$ to mimic its behavior and produce a cosmology in 
agreement with observations.

In the case  $\Lambda=0$, relations (\ref{rhoDE1}), (\ref{pDE1}) become
\begin{eqnarray}
&& \!\!\!\!\!\!\!\!\!\!\!
\rho_{DE}=\frac{3}{8\pi G}H^2\left[1-\alpha \frac{ \delta}{ 2-\delta} 
H^{2(1-\delta) }
\right]
\label{rhoDE3}\\
&& \!\!\!\!\!\!\!\!\!\!\!
p_{DE}= -\frac{1}{8\pi G}\left\{
3H^2\left[1-\alpha\frac{\delta}{2-\delta}H^{
2(1-\delta)}
\right]
\right.
\nonumber\\
&&\left.\ \ \ \ \ \ \ \ \ \ \ \ \ \ \ 
 +2\dot{H}\left[1-\alpha\delta H^{2(1-\delta)}
\right]
\right\},
\label{pDE3}
\end{eqnarray}
while  (\ref{omegaDE}) reads 
\be
 \label{omegaDE1}
\Omega_{DE}(z)=1-\left\{\frac{(2-\delta)}{\alpha\delta} \left[\Omega_{m0}H_0^2 
(1+z)^3\right]^{\delta -1} \right\}^{\frac{1}{\delta-2}}.
\ee 
However, the important simplification comes from expression (\ref{lambda}), that relates 
$\Lambda$ and $\alpha$ with the observationally 
determined quantities $\Omega_{m0}$ and $H_0$. In particular, setting  $\Lambda =0$ 
leads to the determination of parameter $\alpha$ in terms of $\Omega_{m0}$ and $H_0$, 
namely 
\be \label{alpha}
\alpha=\frac{(2-\delta)}{\delta}\Omega_{m0}H^{2(\delta -1)}_{0},
\ee
leaving $\delta$ as the only free model parameter. Note that since $\tilde{\alpha}>0$ in 
(\ref{Tsalsent}), i.e $\alpha>0$, from (\ref{alpha}) we deduce that   the present 
scenario is realized for $\delta<2$.
Thus, inserting (\ref{alpha}) into 
(\ref{omegaDE1}) leads to the simplified expression 
\be
 \label{omegaDE1b}
\Omega_{DE}(z)=1-\Omega_{m0}(1+z)^{\frac{3(\delta-1)}{\delta-2}}.
\ee 
Finally, inserting  (\ref{alpha}) and  (\ref{omegaDE1b}) into (\ref{wDEfinal}) and 
(\ref{qpar}) gives respectively
\be \label{wDEfinal1}
w_{DE}(z)=\frac{(\delta-1)}{(2-\delta)} 
 \left[1-\Omega_{m0}(1+z)^\frac{3(\delta 
-1)}{(\delta -
2)} 
\right]^{-1},
\ee
and
\be \label{qpar1}
q(z)=\frac{2\delta-1}{2(2-\delta)}.
\ee
We stress here that in this case exact $\Lambda$CDM cosmology cannot be obtained for any 
parameter values, and thus one should suitably choose $\delta$ in order to acquire 
agreement with observations. Note that in the standard extensive choice $\delta=1$ we 
obtain a trivial universe with $\Omega_{DE}(z)=1-\Omega_{m0}=const.$ and $w_{DE}(z)=0$.
 
From the analytical expression  (\ref{omegaDE1b}) we can see that we acquire the thermal 
history of the universe, with the sequence of 
matter and dark energy epochs and the onset of late-time acceleration. Furthermore, in 
the future ($z \rightarrow -1$) the universe tends asymptotically to the complete 
dark-energy domination. Additionally, as can be seen from expression  (\ref{wDEfinal1}), 
the asymptotic value of $w_{DE}$ in the far future is not necessarily the cosmological 
constant value $-1$. In particular, we deduce that for 
$1\leq\delta<2$  $w_{DE}\rightarrow0$ as $z \rightarrow -1$, while for 
$\delta<1$    $w_{DE}\rightarrow(\delta-1)/(2-\delta)$ as $z \rightarrow -1$.
Hence, the case $\delta<1$ is the one that exhibits more interesting behavior in 
agreement with observations, and we observe that  for decreasing $\delta$ the $w_{DE}(z)$ 
tends to lower values.

We close this subsection mentioning that according to the above analysis the 
cosmological behavior is very efficient for low redshifts and up to the far future, 
despite the fact that an explicit cosmological constant is absent. However, as can be 
seen from  
(\ref{omegaDE1}), for high redshifts the behavior of  $\Omega_{DE}(z)$ is not 
satisfactory, since as it is this expression leads to either early-time dark energy or to 
the unphysical result that  $\Omega_{DE}(z)$ becomes negative. In order to eliminate this 
behavior and obtain a universe evolution in agreement with observations at all redshifts 
one needs to include the radiation sector too, which indeed can regulate the early-time 
behavior. This is performed in the next subsection.

 \subsection{Cosmological evolution including radiation}
 
 In this subsection for completeness we extend the  scenario of modified cosmology 
through 
nonextensive horizon thermodynamics, in the case where the radiation fluid is also 
present. 
First of all, in the case where extra fluids are considered in the universe content, the 
thermodynamical procedure of Section \ref{Themodel} is applicable in exactly the same 
way, with the only straightforward addition being that in Eq. (\ref{energy}) one should 
add the energy densities and pressures of all   universe fluids
\cite{Padmanabhan:2009vy,Frolov:2002va,Cai:2005ra,Akbar:2006kj,Izquierdo:2005ku,
Jamil:2010di}. Hence, if we allow for a radiation fluid, with energy density $\rho_r$ and 
pressure $p_r$, and repeat the analysis of subsection \ref{fullmodelconstr}, the 
Friedmann equations (\ref{FR1}), (\ref{FR2}) become
\begin{eqnarray}
\label{FR1rad}
&&
\!\!\!\!\!\!\!\!\!\!
H^2=\frac{8\pi G}{3}\left(\rho_m+\rho_r+\rho_{DE}\right)\\
&&
\!\!\!\!\!\!\!\!\!\!
\dot{H}=-4\pi G \left(\rho_m+p_m+\rho_r+p_r+\rho_{DE}+p_{DE}\right),
\label{FR2rad}
\end{eqnarray}
with  $\rho_{DE}$, \ $p_{DE}$ still given by (\ref{rhoDE1}), (\ref{pDE1}), and $w_{DE}$ 
by 
(\ref{wDE}).
 
 We proceed by introducing the radiation density parameter as
\begin{eqnarray}
  \Omega_r\equiv\frac{8\pi G}{3H^2}\rho_r,
 \label{Omrad}
 \end{eqnarray} 
 and thus the first Friedmann equation becomes   $\Omega_r+ \Omega_m+\Omega_{DE}=1$.
Similarly to the analysis of section \ref{CosmEvol}, in order to extract analytical 
expressions we consider that the matter fluid is dust, namely $w_m=0$. In the case where 
radiation is present we still have that $\Omega_m=\Omega_
{m0} H_{0} ^2/a^3 H^2$, however (\ref{h2}) now extends to
\be \label{h2rad}
H=\frac{\sqrt{\Omega_{m0}} H_{0}}{\sqrt{a^3 (1-\Omega_{DE}-\Omega_{r})}},
\ee
while (\ref{hddot}) reads
\be \label{hddotrad}
\dot 
H=-\frac{H^2}{2}\left[\frac{3\Omega_{m0}+4\Omega_{r0}(1+z)}{\Omega_{m0}
+\Omega_{
r0}(1+z)}+\frac{(1+z)\Omega_{DE}'}{(1-\Omega_{DE})}\right],
\ee
since for dust matter we have
\be 
\label{omegarrad}
\Omega_{r}(z)=\frac{\Omega_{r0}(1+z)^{4}[1-\Omega_{DE}(z)]}{\Omega_{m0}
(1+z)^{3}+\Omega_{ r0 } (1+z)^{4}}.
\ee

\subsubsection{Cosmological evolution with $\Lambda\neq 0$}

Let us first investigate the case where  $\Lambda\neq 0$. Inserting (\ref{rhoDE1}) into 
(\ref{ode}) and using (\ref{h2rad}) we find that (\ref{omegaDE})  extends to
 \begin{eqnarray} 
 \label{omegaDErad}
&&
\!\!\!\!\!\!\!\!\!\!\!\!\!\!\!\!
\Omega_{DE}(z)=
1-H^{2}_{0}\left[\Omega_{m0}(1+z)^3+\Omega_{r0}(1+z)^4\right]\nonumber\\
&&\!\!\!\!\!\!\!\!\!\!\!\!\!\!\!
\cdot\left\{\frac{
(2\!-\!\delta)}{\alpha 
\delta}\!\left[H^{2}_{0}\left[\Omega_{m0}(1\!+\!z)^3\!+\!\Omega_{r0}(1\!+\!z)^4\right]
\!+\!\frac { \Lambda }{ 3} \right]\right\}^{\frac{1}{\delta -2}}\!.
 \end{eqnarray}
 This expression is the analytical solution for the dark energy density parameter  
$\Omega_{DE}(z)$, in a flat universe and for dust matter, in the case where radiation is 
present. Applying it at present time, 
i.e at $z=0$, we acquire
\begin{equation} \label{lambdarad}
 \Lambda=\frac{3\alpha 
\delta}{2-\delta}H^{2(2-\delta)}_{0}-3H^{2}_{0}\left(\Omega_{m0}+\Omega_{r0}\right),
\end{equation}
 which provides the relation that relates $\Lambda$, $\delta$ and $\alpha$ with the
observationally 
determined quantities $\Omega_{m0}$, \ \ $\Omega_{r0}$ and $H_0$, leaving the scenario 
with 
two free 
parameters. As expected, for $\delta=1$ and $\alpha=1$
all the above relations give those of $\Lambda$CDM cosmology with radiation sector 
present.

Differentiating (\ref{omegaDErad}) we find
 \begin{eqnarray} 
\label{omegaDEdotrad}
&&
\!\!\!
\Omega'_{DE}(z)=\mathcal{A}(z)\mathcal{B}^{\frac{1}{\delta -2}}(z)\left[\frac{1}{\alpha 
\delta}\mathcal{B}^{-1}(z)\mathcal{C} (z)-1\right],
 \end{eqnarray}
 where $
\mathcal{A}(z)=H^{2}_{0}\left[3\Omega_{m0}(1+z)^2+4\Omega_{r0}(1+z)^3\right]
$, 
$
\mathcal{B}(z)=\frac{2-\delta}{\alpha 
\delta}\left[H^{2}_{0}\left[\Omega_{m0}(1+z)^3+\Omega_{r0}(
1+z)^4\right]+\frac{\Lambda}{3}\right]
$ and 
$\mathcal{C}(z)=H^{2}_{0}\left[\Omega_{m0}(1+z)^3+\Omega_{r0}(1+z)^4\right]
$.
 Hence,  $w_{DE}(z)$ is calculated from  (\ref{wDE}), but now eliminating $\dot{H}$ 
through 
(\ref{hddotrad}), obtaining
\begin{widetext}
\be 
\label{wDEfinalrad}
w_{DE}(z)=-1+\frac{\left[\frac{3\Omega_{m0}+4\Omega_{r0}(1+z)}{\Omega_{m0}+\Omega_{
r0}(1+z)}(1-\Omega_{
DE})+(1+z)\Omega_{DE}'\right]\left\{1-\alpha \delta 
\left[\frac{H^{2}_{0}\left[\Omega_{m0}(1+z)^3+\Omega_{r0}(1+z)^4\right]}{(1-\Omega_{DE})}
\right]^{1-\delta}\right\}}{(1-\Omega_{DE} )\left\{\frac{\Lambda 
(1-\Omega_{DE})}{H^{2}_{0}\left[\Omega_{m0}(1+z)^3+\Omega_{r0}(1+z)^4\right]}+3\left\{
1-\frac{\alpha 
\delta}{2-\delta}\left\{\frac{H^{2}_{0}\left[\Omega_{m0}(1+z)^3+\Omega_{r0}(1+z)^4\right]}
{(1-\Omega_{DE})}\right\}^{1-\delta}\right\}\right\}},
\ee
\end{widetext}
  where $\Omega_{DE}$ and $\Omega_{DE}'$ are given by (\ref{omegaDErad}) and 
(\ref{omegaDEdotrad}) respectively. 
Lastly,   the decelaration parameter $q\equiv 
-1-\frac{\dot H}{H^2}$, using  (\ref{hddotrad}) is found to be 
\be \label{qparrad}
q(z)=-1+\frac{1}{2}\left[\frac{3\Omega_{m0}+4\Omega_{r0}(1+z)}{\Omega_{m0}
+\Omega_{
r0}(1+z)}+\frac{(1+z)\Omega_{DE}'}{(1-\Omega_{DE})}\right].
\ee

In summary, in the case where radiation is present, we were able to extract analytical 
solutions for $\Omega_{DE}(z)$ and $w_{DE}(z)$, for the modified, nonextensive 
cosmological scenarios of the present work.

\subsubsection{Cosmological evolution with $\Lambda= 0$}

Let us now focus on the interesting case where the explicit 
cosmological constant is absent, namely when $\Lambda=0$. This  scenario was analyzed in 
subsection \ref{Lam0a} above in the absence of radiation, however we now study it in the 
full case where radiation is included. For $\Lambda=0$, relation  (\ref{omegaDErad}) 
becomes
 \be
 \label{omegaDEradbb}
\Omega_{DE}(z)=1-\left\{\frac{\left[\Omega_{m0}(1+z)^{3}+\Omega_{r0}(1+z)^{4}\right]^{
\delta 
-1}}{\Omega_{m0}+\Omega_{r0}}\right\}^{\frac{1}{\delta -2}},
\ee
relation (\ref{lambdarad})
 becomes 
\be
 \label{lambdaradbb}
\alpha=\frac{2-\delta}{\delta}H^{2(\delta-1)}_{0}\left[\Omega_{m0}+\Omega_{r0}\right],
\ee
and thus positivity of $\alpha$ implies that $\delta<2$,
relation (\ref{omegaDEdotrad})
 becomes
\begin{eqnarray}
 \label{omegaDEdotradbb}
&&
\!\!\!\!\!
\Omega_{DE}'(z)=-\frac{\delta -1}{\delta 
-2}\left\{\frac{\left[\Omega_{m0}(1+z)^3+\Omega_{r0}(1+z)^4\right]}{\left[\Omega_{m0}
+\Omega_{r0}\right]}\right\}^{\frac{1}{\delta -2}}\nonumber\\
&&
\ \ \ \ \ \ \ \ \ \ \ \ \ \ \ \  \ \ 
\cdot
\left[3\Omega_{m0}(1+z)
^2+4\Omega_{r0}(1+z)^3\right],
\end{eqnarray}
relation (\ref{wDEfinalrad}) becomes 
\begin{widetext}
\begin{eqnarray}
 \label{wDEfinalradbb}
&&\!\!\!\!\!\!\!\!\!\!
w_{DE}(z)=\frac{3(1-\delta)\Omega_{m0}+(2-3\delta)(1+z)\Omega_{r0} }{3(\delta-2) 
[\Omega_{m0}+(1+z)\Omega_{r0}]} \nonumber\\
&&
\ \ \ \ \ \ \ \ \ \ 
+
\frac{(\delta-1) [3\Omega_{m0}+4(1+z)\Omega_{r0}]}
{3(\delta-2) [\Omega_{m0}+(1+z)\Omega_{r0}]- 3(\delta-2)
(1+z)^{\frac{3(1-\delta)}{\delta-2}}(\Omega_{m0}+\Omega_{r0})^{\frac{1}{\delta-2}}[\Omega_
{m0}+(1+z)\Omega_{r0}]^{\frac{1}{2-\delta}}}     ,
\end{eqnarray}
\end{widetext}
while relation (\ref{qparrad}) for the deceleration parameter $q$ becomes \\

\begin{eqnarray}
 \label{qparradbb}
&&\!\!\!\!\!\!\!\!\!\!\!\!\! \!\!\!
 q=\frac{\left[\Omega_{m0}+2\Omega_{r0}(1+z)\right]}{2\left[\Omega_{m0}+\Omega_{
r0} (1+z)\right]}
\nonumber\\
&&\!\!\!\!\!
 -\frac{\delta -1}{2(\delta 
-2)}\left\{\frac{\left[3\Omega_{m0}(1+z)^3+4\Omega_{r0}(1+z)^4\right]}{\left[\Omega_{m0}
(1+z)^3+\Omega_{r0}(1+z)^4\right]}\right\}.
\end{eqnarray}
We mention that relations (\ref{omegaDEradbb})-(\ref{qparradbb}) are the extensions of 
(\ref{omegaDE1})-(\ref{qpar1}) in the presence of radiation.

Let us examine this scenario in more detail, and in particular   study the effect of 
$\delta$ on the cosmological evolution. In Fig. \ref{fig:fig4} we present $w_{DE}(z)$ 
for various choices of $\delta$, extending the evolution up to the far future. In all 
cases the parameter $\alpha$ is set according to  (\ref{lambdaradbb}) in 
order to obtain $\Omega_{m}(z=0)=\Omega_{m0}=0.3$ and 
$\Omega_{r}(z=0)=\Omega_{r0}=0.000092$ \cite{Ade:2015xua}, and the expected thermal 
history of the universe. As we observe, for decreasing $\delta$ the $w_{DE}(z)$ tends to 
lower values. Moreover, although the asymptotic value of $\Omega_{DE}(z)$ as  $z 
\rightarrow -1$ is 1, as can be seen immediately from (\ref{omegaDEradbb}), namely the 
universe tends to the complete dark-energy 
domination, the asymptotic value of  $w_{DE}$ is not the cosmological 
constant value $-1$, i.e the universe does 
not   result in a de Sitter space. In particular, from (\ref{wDEfinalradbb}) we can 
see that for $1\leq\delta<2$, $w_{DE}\rightarrow0$ as $z \rightarrow -1$,  while for 
$\delta<1$,  $w_{DE}\rightarrow(\delta-1)/(2-\delta)$ as $z \rightarrow -1$. 
These asymptotic values are the same with the ones in the absent of radiation mentioned 
in subsection  \ref{Lam0a}, which was expected since at late times the effect of 
radiation is negligible.
\begin{figure}[!h]
\centering
\includegraphics[width=8.2cm]{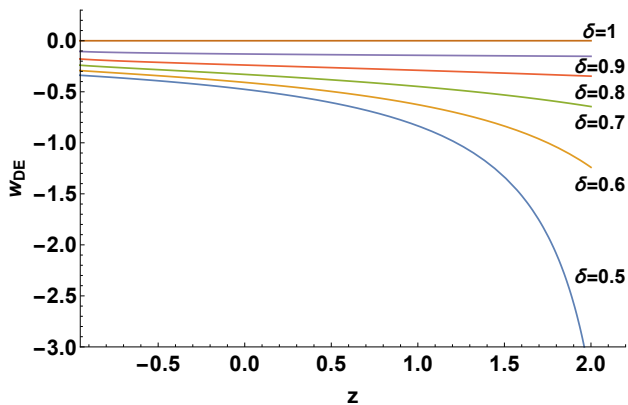}
\caption{\it{The evolution of the equation-of-state parameter $w_{DE}$ as a function of 
the redshift $z$, for  $\Lambda=0$ and for various values of the nonextensive parameter 
$\delta$, in the case where radiation is present. For each value of $\delta$ we  choose  
$\alpha$ according to (\ref{lambdaradbb})  in order to obtain  
$\Omega_{m}(z=0)=\Omega_{m0}=0.3$ and 
$\Omega_{r}(z=0)=\Omega_{r0}=0.000092$ at present \cite{Ade:2015xua}, and acquire the 
expected 
thermal history of the universe.}}
\label{fig:fig4}
\end{figure}
 
In summary, the scenario of  modified cosmology through nonextensive thermodynamics, even 
in the case where an explicit cosmological constant is absent, is efficient in describing 
the cosmological behavior of the universe. In order to present this behavior more 
transparently we confront the scenario  with Supernovae type Ia (SN Ia) data. In these 
observational sets  the apparent luminosity  $l(z)$, or equivalently the apparent 
magnitude $m(z)$, are measured as functions of the
redshift, and are related to the luminosity distance as
\begin{equation}
2.5 \log\left[\frac{L}{l(z)}\right] = \mu \equiv m(z) - M = 5 
\log\left[\frac{d_L(z)_{\text{obs}}}{Mpc}\right]  + 25,
\end{equation}
where $M$ and $L$ are the absolute magnitude and luminosity respectively.
Additionally, for any theoretical model one can calculate the predicted dimensionless 
luminosity distance $d_{L}(z)_\text{th}$ using the predicted evolution of the Hubble 
function   as
\begin{equation}
d_{L}\left(z\right)_\text{th}\equiv\left(1+z\right)
\int^{z}_{0}\frac{dz'}{H\left(z'\right)}~.
\end{equation}
In the scenario at hand, $H(z)$ can be immediately calculated analytically from 
(\ref{h2rad}), knowing (\ref{omegarrad}) and (\ref{omegaDErad}).
In Fig.   \ref{SNd} we depict the theoretically predicted apparent minus absolute 
magnitude  as a function of $z$, for two $\delta$ choices, as well as the prediction of 
$\Lambda$CDM cosmology, on top of the   $580$ SN Ia observational data points from 
\cite{Suzuki:2011hu}. As we can see the agreement with the SN Ia data is excellent. 
The detailed comparison with observations, namely the joint analysis using 
data from SN Ia,   Baryon Acoustic Oscillation (BAO), Cosmic Microwave Background (CMB), 
and direct   Hubble parameter observations, lies beyond the scope of the 
present work and it is left for a future project.
  \begin{figure}[ht]
\includegraphics[scale=0.64]{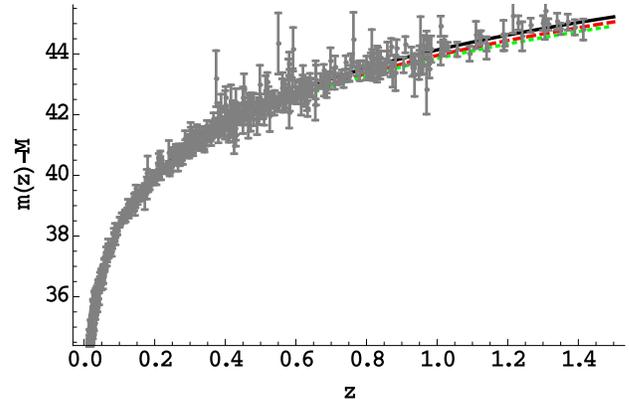}
\caption{
{\it{The theoretically predicted apparent minus absolute magnitude as a function of the 
redshift, for  the scenario of  modified cosmology through nonextensive thermodynamics,
for $\Lambda=0$, in the case where radiation is present,  for 
$\delta=0.5$ (red-dashed) and 
$\delta=0.6$ (green-dotted). The observational points correspond to the  $580$ SN Ia data 
points from
\cite{Suzuki:2011hu}, and for completeness and comparison  we 
depict the prediction of  $\Lambda$CDM cosmology with the black-solid curve.
}} }
\label{SNd}
\end{figure}

We close this subsection mentioning that the present scenario is very efficient in  
mimicking the cosmological constant, despite the fact that in this case the exact 
$\Lambda$CDM cosmology cannot be obtained for any parameter values. In particular, 
choosing the nonextensive parameter $\delta$ suitably (namely  $\delta\sim0.5-0.6$) we 
acquire agreement with observations. This is a significant result that shows the 
capabilities of the modified cosmology through nonextensive thermodynamics.

\section{Conclusions}
\label{Conclusion}

In this work we constructed a modified cosmological scenario through the application of 
the first law of thermodynamics, but using the generalized, nonextensive Tsallis entropy 
instead of the usual Bekenstein-Hawking one. In particular, there is a well-studied 
procedure in the literature, which works for a variety of modified gravities,
where one can apply the first law of thermodynamics in the universe horizon and extract 
the Friedmann equations. The crucial part in this procedure is the use of the modified 
entropy relation of the specific modified gravity, which is known only after this 
modified gravity is given, and thus in this sense it cannot provide new gravitational 
modifications. However, if we apply this approach  using the nonextensive, Tsallis 
entropy, which is the consistent concept that should be used in non-additive 
gravitational systems such us the whole universe, then we result to modified cosmological 
equations that possess the usual ones as a particular limit, but which in the general case 
contain extra terms that appear for the first time. 

The new terms that appear in the modified Friedmann equations are quantified by the 
nonextensive parameter $\delta$ and constitute an effective dark energy sector. In the 
case where Tsallis entropy becomes the usual Bekenstein-Hawking entropy, namely when 
$\delta=1$, the effective dark energy coincides with the cosmological constant and 
$\Lambda$CDM cosmology is restored.  However, in the general case the scenario of 
modified cosmology at hand presents very interesting cosmological behavior. 

When the matter sector is dust, we were able to extract analytical expressions for the 
dark energy density and equation-of-state parameters, and we extended these solutions in 
the case where radiation is present too. These solutions show that the universe exhibits 
the usual thermal history, with the sequence of matter and dark-energy eras and the onset 
of acceleration at around $z\approx0.5$ in agreement with observations. In the case where 
an explicit cosmological constant is present, according to the value of $\delta$ the 
dark-energy equation-of-state parameter exhibits a very interesting behavior and it can 
be 
quintessence-like, phantom-like, or experience the phantom-divide crossing during the 
evolution, before it asymptotically stabilizes in the cosmological constant value $-1$ in 
the far future.

An interesting sub-case of the scenario of modified cosmology through nonextensive 
thermodynamics is when we set the explicit cosmological constant to zero, since in this 
case the universe evolution is driven solely by the news terms. Extracting analytical 
solutions for the dark energy density and equation-of-state parameters we showed that 
indeed the new terms can very efficiently mimic $\Lambda$CDM cosmology, although 
$\Lambda$ is absent, with the 
successive  sequence of matter and dark energy epochs, before the universe results in 
complete dark-energy domination in the far future. Moreover, confronting the model with 
SN Ia data we saw that the agreement is excellent. 

In summary,  modified cosmology through nonextensive thermodynamics is very efficient in 
describing the universe evolution, and thus it can be a candidate for the description of 
nature. In the present work we derived the cosmological equations by applying the 
well-known thermodynamics procedure to the universe horizon. It would be interesting to 
investigate whether these equations can arise from a nonextensive action too. Such a 
study is left for a future project.

\section*{Acknowledgments}
 
This article is based upon work from CANTATA COST (European Cooperation in Science and 
Technology) 
action CA15117, EU Framework Programme Horizon 2020.

\end{document}